\documentstyle[12pt,prl,aps,graphicx]{revtex}

\begin{document}
 \title{ \Large 
Surprises of Phase Transition Astrophysics}
\normalsize \author{  Zakir F. Seidov, 
Dept of Physics, POB 653 Ben-Gurion University, \\84105 Beer-Sheva, Israel 
\\
E-mail: seidov@bgumail.bgu.ac.il}
\maketitle 
\begin{abstract} 

It is a half-century-long story about the first-order phase transition
effects
on equilibrium, stability and pulsations of planets and stars. The
topics more or less considered in author's papers are mainly touched, and
no attempt was done to cover all literature on subject.\\
It was Ramsey who in 1950 had first shown in MNRAS paper [1] that $if$
in the center
of a planet, with central pressure $P_c\;$ just reaching critical
pressure $P_0$, the first-order phase transition takes place with density
jump from $\rho_1$ to $\rho_2=q*\rho_1$, and $if$ $q>1.5$, then planet
loses its stability {\em at the same moment}.\\
 This remarkably amazing and
quite unexpected result has been since then rediscovered by many authors
repeatedly - last time in nucl-th/9902033.\end{abstract}
  \begin{flushright}                  
     {\em   "...- D'you think  all these (cloths) will be the wear?}\\
            {\em       - I think all  these should be tailored."}\\
                          Yuri Levitanski, Soviet Jewish poet
  \end{flushright}                  

The story  began to me some 35 years ago when trying to understand
paper [2], 
I've found that something is wrong with Mass-Central Density curves
for white dwarf stars near maximum - there should be
$sharp$, not $smooth$ maximum of mass, as reverse $\beta$-decay reactions
lead to {\em the discontinuity of density distribution}  inside the
star.

To simplify the problem, I used some models, allowing
$analytical$ investigation - \mbox{polytropes} with indices $n$ equal to
$0$
and $1$ in envelope and core in various combinations. In all cases
considered, it happened
that $if$ $q>1.5$, the instability occurs at the moment the
phase transition starts in the center of the star.

My supervisor acad. Ya.B. Zeldovich for a long time did not believe in
such
$"strange"$
figure $3/2$, so I kept analyzing various models. After a while
 Ya.B. became himself convinced and within a brief period he developed
an amazingly elegant
method of prooving this {\em  word  constant  3/2}, and a joint
paper [4] was submitted to Astronom. Zhurnal. Needless to say, I was happy
that I
was right and
that I had joint paper with such an
outstanding scientist. Catastrophe came short: Ya.B. sent me a brief
letter noting that {\em "the result 3/2 is known in literature!"}.

 No comments...

 I've never managed since then to write another joint paper with
Ya.B....

Why $3/2$? I do not know, but
it seems that $3/2\; number$ is
directly related to the \mbox{$r^{-1}$ - law} of potential $U(r)$
in Newtonian Theory of Gravitation (NTG) in $3D$ space[11]. 

This is not the last surprise of Phase-Transition Astrophysics (PTA)!

In the classical theory [5] of equilibrium and stability of stars
comprised of matter
with $smooth$ Equations of State (EoS), there is a common sense that
{\em rotation leads to the increase of stability reserve} while
{\em General Relativity (GR) leads to the decrease of stability}. 

And you may guess that exactly $opposite$ situation is in PTA.
In GR, the critical value of $energy$ density
$q=\varepsilon_2/\varepsilon_1$
is [6]: $3/2\;(1+P_0/\varepsilon_1),$ 
that is $larger$ than in NTG.

Why $larger?$ - I do not know\dots 

As to rotation, it was found [7], that for steady-state rotation with
$small$ angular velocity $\Omega$, 
$q_{crit}=3/2\;-\;\Omega^2\;/\;4\pi G \rho_1$, 
that is, of course, rotation $reduces$ the stability
of star $against$ phase-transition-induced instability.

Why $reduces?$ - I do not know...

Abovementioned {\em three suprises} of PTA, combined in the formula
$$q_{crit}=3/2\;-\;\Omega^2\;/\;4\pi G \rho_1\;
 + 3/2\;(1+P_0/\rho_1 c^2),$$ 
are valid for $any$ EoS's of old and new phases.

A number of amazing results was found analysing various particular
models. First to be mentioned is the two-constant-density-phase model with 
First-Order Phase Transition (PT1) at the boundary between core and
envelope. This model was extensively used
for different problems - general dependence of Mass-Radius etc. curves,
effects of rotation and GR, neutral core (when PT1 {\em begins at some
distance from the center of a star}) etc.

In the last case, it was found~[3] 
that $q_{crit}$ is an increasing function of size of neutral core:
$q_{crit}=1+k/[3-4\;x-(k-1)\;x^4],$ where $x$ is $relative$ radius 
of neutral core, and $k$ is relation of density in neutral core to
density in envelope; and $q_{crit}\rightarrow \infty$ at  
$x\;\rightarrow x_{cr}(k)$, where for example at $k=1$, $x_{cr}=3/4$ -
another amazing value. At $larger$ neutral cores, PT1 with arbitrary large
$q$ can not force a star to lose its dignity and stability!

 For a model with polytropic indices $n=1$ both in envelope and neutral
core~[3], $x_{cr}=.6824$ (corresponding relative mass of core is .6375).

Returning to $n=0$, for $q>3/2$ there is another critical point in
$Mass-P_c$ curves, where at the minimum of mass the $recovery$ of stability
takes place, and for larger $P_c$ there is a branch of stable
equilibrium
states. At $the minimum\;\;of\;\;Mass$:\begin{equation}
f(q,x)=(q-1)^2 x^4\;+\;4\;(q-1)\;x + (3-2 q) =0.\end{equation}
It was found [8] that GR effects, in $the\; first\;
Post-Newtonian\;approximation$
lead to the PN-correction to $x$ in Eq.~(1): \begin{equation}
\Delta_{PN}(q,x)={9-7\;q+27(q-1)x+(q-1)(4q-27)x^2+(q-1)(9-4q)x^3
\over {2(q-1)[1+(q-1)x^3]^3}}{P_0 \over \rho_1 c^2}.\end{equation}
The surprises have not finished - this correction is $negative$ at small
values of $q$ and $positive$ at $q>1.89$. That is for larger $q$, GR
effect is of $correct$ sign and $reduces$ the region of stability
at $(q-x)$ plane.

For the same $n=0$ model, the analytical formula for
frequency, $\omega$, of small adiabatic radial pulsations of the 
lowest mode can be found:
\begin{equation}
\omega_0^2={{4\pi\;G\;\rho_1\;f(q,x)}\over {3\;(q-1)(1-x)}},\end{equation}
and in the case of  $slow$ rotation with angular
velocity $\Omega\;$[10]:
$$\omega_{\Omega}^2=\omega_0^2+
\Delta_{\Omega}(q,x),\;\;\;\mbox{with}$$
\begin{equation}
\Delta_{\Omega}(q,x)={\frac 23}\; \Omega^2   
\left[\frac{5x(1-x)(1+x)^2}{1+(q-1)x^5}-
\frac {1+(q-1)x}{(q-1)(1-x)}\right].\end{equation} 

 Rotational correction to frequency squared is $negative$ at smaller
values of $x$, that is rotation reduces the stability reserve of star with
PT1, then for larger value of $x$, rotational effect is of $"correct"$ sign.

In fact, dependence of both corrections, due to GR and rotation, on $x$
and $q$ is rather complicated, and Fig.~1 presents only a part of
$(q-x)\--$plane
with lines on which $\Delta_{\Omega}(q,x)=0$ (broken line labeled as "Rot")and
$\Delta_{PN}(q,x)=0$ (dash-point line labeled as "GR").
Also shown is the curve $f(q,x)=0$ (solid line labeled as "Crit") from Eq.1,  
which marks a boundary between $stable$ equilibrium states (right-hand) 
from $unstable$ ones (left-hand).  Remarkably, this curve
crosses {\em both} curves of $zero\;correction$. Due to GR, curve 
$f(q,x)=0$ is forced to
rotate clockwise around point of intersection of lines "GR" and "Crit", while rotation
makes the critical curve rotate counterclockwise around point of intersection of lines 
"Rot" and "Crit".
\begin{figure}
\includegraphics[scale=.5]{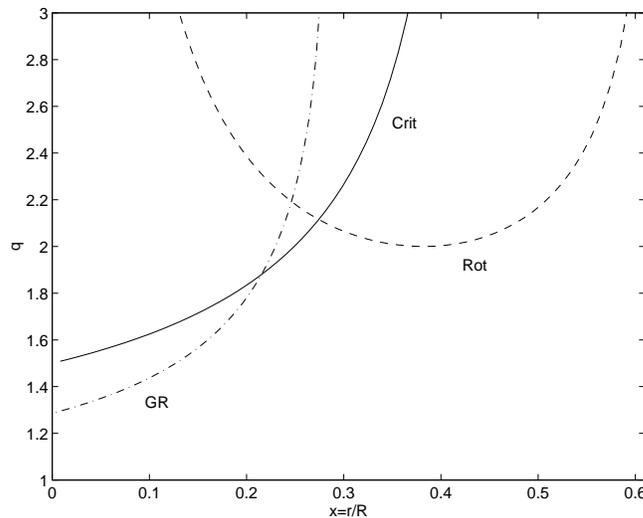}
\caption{Three important curves at $(q-x)\--$plane: critical state of
star stability
recovery, see Eq.~1, and GR and Rotational corrections equal to zero,
see Eq.~2 and Eq.~4, respectively.}
\end{figure}
Why do rotation and GR exert such strange effects on stability of a star
with PT1?

- {\em I do not know\dots}

A lot of interesting and unsolved problems of
pulsations (eigen- values and functions, damping), strong rotation with
regard to deviation of equilibrium figures from spherical
symmetry, considering more EoS, etc.
have been  left aside , but 1500-word mark is close and
I pass to the epilogue...

Most of this story happened to me many years ago, in the Soviet Union,
the Power so unexpectedly crushed in the recent past, as if some
first-order phase transition had acted in the center of It...

Now I'm in Israel, all my papers being left over there and these lines
being written by heart, by memory, with no paper at hand...

And last sentences (hopefully still in 1500-word limit), returning to
epigraph:

- Do I believe  this all {\em will be} awarded?  

- I think  this all {\em should be} written.

\newpage Some bibliographical remarks:

1. W.H. Ramsey, MNRAS {\bf 110} (1950) 325; {\bf 113} (1951) 427;\\
see also
M.J. Lighthill, MNRAS {\bf 110} (1950) 339 ($n=0$ model),\\
W.C. De Markus, Astron. J. {\bf 59} (1954) 116 (rotation).

2. T. Hamada, E.E.Salpeter, Ap.J. {\bf 134} (1961) 669;\\
see also
E. Schatzman, Bull. Acad. Roy. Belgique {\bf 37} (1951) 599;\\
E. Schatzman, White Dwarfs, North-Holland Publ. Co. Amsterdam, 1958.

3. Z.F. Seidov, 
Izv. Akad. Nauk Azerb. SSR, ser. fiz-tekh. matem. \\ no. 5 (1968) 93
(n=0);\\
Soobsh. Shemakha Astrophys. Observ. {\bf 5} (1970) 58 (n=1);\\
Izv. Akad. Nauk Azerb. SSR, ser. fiz-tekh. matem. \\ no. 1-2 (1970) 128
(n=0 with neutral core);\\
Izv. Akad. Nauk Azerb. SSR, ser. fiz-tekh. matem. \\ no. 6 (1969) 79 
(n=1 with neutral core).\\
Ph.D thesis. Yerevan Univ. 1970.

4. Ya.B. Zeldovich, Z.F. Seidov (1966) unpublished;\\see also 
Z.F. Seidov, Astrofizika {\bf 3} (1967) 189 .

5. Ya.B. Zeldovich, I.D. Novikov, Relativistic Astrophysics, Univ. Chicago
Press, Chicago, 1971 (this is only one of many possible references to
these outstanding authors).

6. Z.F. Seidov, Astron. Zh. {\bf 48} (1971) 443 (GR);\\ 
see also B. K{\"a}mpfer, Phys.Lett. {\bf 101B} (1981) 366.

7. Z.F. Seidov, Astrofizika {\bf 6} (1970) 521 (rotation).

8. Z.F. Seidov, Space Research Inte Preprint (1984) Pr-889 (GR).

9. M.A. Grienfeld, Doklady Acad. Nauk SSSR {\bf 262} (1982) 1342
(pulsations);\\
see also G.S. Bisnovatyi-Kogan, Z.F. Seidov, Astrofizika {\bf 21} (1984) 570. 

10. Z.F. Seidov, Doctor of Sci. Theses. Space Research Inte,
Moscow, 1984.

11. Z.F. Seidov, astro-ph/9907136 (non-1/r potential and PT1).
\end{document}